\documentclass{iopjournal}
\usepackage{multirow}

\begin{document}

\articletype{Paper} 

\title{Are Virtual DES Images a Valid Alternative to the Real Ones?}

\author{Ana C. Perre$^1$\orcid{0000-0001-6668-2620}, Lu\'{i}s A. Alexandre$^2$\orcid{0000-0002-5133-5025} and Lu\'{i}s C. Freire$^{3}$\orcid{0000-0000-0000-0000}}

\affil{$^1$Faculdade Ciências da Saúde, Universidade da Beira Interior and Unidade Local de Saúde do Oeste, Av. Infante D. Henrique, 6200-506, Covilhã, Portugal}

\affil{$^2$NOVA LINCS, Universidade da Beira Interior, Rua Marquês d'Ávila e Bolama, 6201-001, Covilhã, Portugal}

\affil{$^3$Escola Superior de Tecnologia da Saúde de Lisboa,  Instituto Politécnico de Lisboa, Av. D. João II, lote 4.69.01, Parque das Nações, 1990-096, Lisboa, Portugal}

\email{ana.perre@ipcb.pt}



\begin{abstract}
\textbf{Contrast-enhanced spectral mammography (CESM) is an imaging modality that provides two types of images, commonly known as low-energy (LE) and dual-energy subtracted (DES) images. In many domains, particularly in medicine, the emergence of image-to-image translation techniques has enabled the artificial generation of images using other images as input. Within CESM, applying such techniques to generate DES images from LE images could be highly beneficial, potentially reducing patient exposure to radiation associated with high-energy image acquisition.
In this study, we investigated three models for the artificial generation of DES images (virtual DES): a pre-trained U-Net model, a U-Net trained end-to-end model, and a CycleGAN model. We also performed a series of experiments to assess the impact of using virtual DES images on the classification of CESM examinations into malignant and non-malignant categories.
To our knowledge, this is the first study to evaluate the impact of virtual DES images on CESM lesion classification. The results demonstrate that the best performance was achieved with the pre-trained U-Net model, yielding an F1 score of 85.59\% when using the virtual DES images, compared to 90.35\% with the real DES images. This discrepancy likely results from the additional diagnostic information in real DES images, which contributes to a higher classification accuracy. 
Nevertheless, the potential for virtual DES image generation is considerable and future advancements may narrow this performance gap to a level where exclusive reliance on virtual DES images becomes clinically viable.}
\end{abstract}

\section{Introduction}
\label{sec:Intro}
Digital mammography (DM) is often considered the gold standard imaging modality in breast cancer diagnosis; however, it has certain limitations, particularly in dense breasts, where tissue overlap may occur, leading to a significant number of inconclusive lesions. \cite{Khaled2022,Song2022}. 
In 2011, the Food and Drug Administration (FDA) granted approval for Contrast-enhanced Spectral Mammography (CESM) as a complementary imaging modality to DM and ultrasound examinations. This imaging modality effectively overcomes the limitations associated with tissue overlap and offers information comparable to that obtained from magnetic resonance imaging (MRI), facilitating the location and characterization of occult or inconclusive lesions. \cite{Khaled2022,Song2022}.

\begin{figure}
 \centering
  \includegraphics[width=1\textwidth]{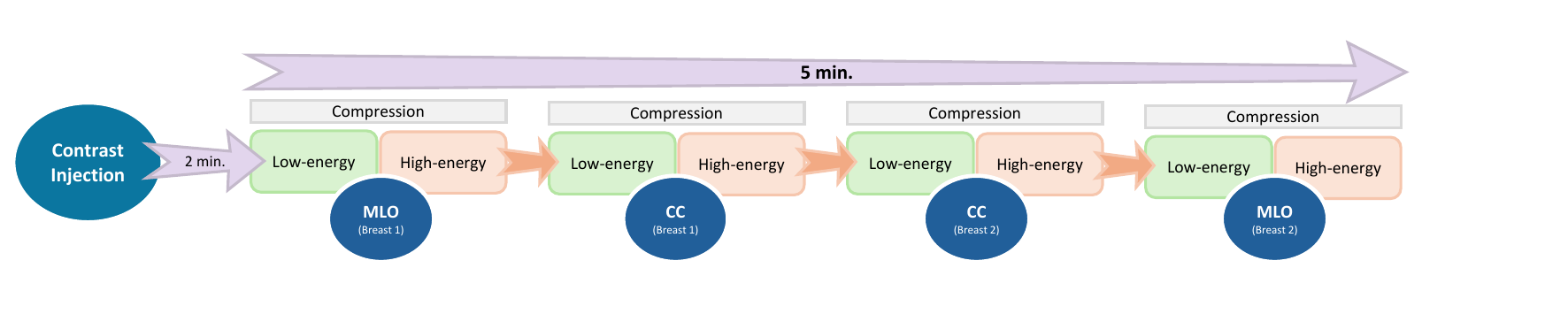}
 \caption{Example of CESM examination: starts with breast 1, the suspicious one, followed by breast 2, with no suspicion of pathology.}
\label{fig:CESM}
\end{figure}

Generally, in a CESM examination, low- and high-energy images are sequentially acquired 2 minutes after intravenous injection of iodated contrast media (1.5 ml/kg), for each view of the breast (Cranio-Caudal - CC - and Mediolateral Oblique - MLO) \cite{Perek2019,Khaled2022} - see Fig. \ref{fig:CESM}. 
The outputs are therefore the contrast-enhanced low-energy (LE) and the dual-energy subtraction (DES) images \cite{Danala2018}. 
These images often allow the identification of breast regions with increased or leaky vascularity, which is common in neoplasms \cite{Patel2018,Sensakovic2021}.  

Since high-energy images are only used to generate DES images \cite{Khaled2022}, their acquisition could be avoided if the DES images were artificially generated from the corresponding LE images. This procedure would offer several advantages, not only in terms of workflow, but also in the reduction of the radiation dose to which patients are subjected, potentially reducing radiation exposure in the CESM examination. This work aims to assess if such virtual DES images can be a valid replacement for the real DES images when it comes to training a classification architecture to differentiate malignant and non-malignant breast images.

Several papers have demonstrated that the image-to-image translation technique has been effectively utilized across different imaging modalities, such as Magnetic Resonance Imaging (MRI), Computed Tomography (CT), and Mammography, allowing efficient cross-modality translation that improves diagnostic accuracy and analytical capabilities \cite{Khorshidifar, 2024.ROFENA.1}. 
In the field of CESM, this technique is beginning to be explored. Since 2024, a variety of models have been used to artificially create DES images, including Autoencoder, Pix2Pix, U-net, Resnet-18 and CycleGAN \cite{2024.ROFENA.1, Khorshidifar, Hosseinipour}. 
Until now, there has been a consensus that CycleGAN is the model that provides the best quantitative results when evaluating Peak Signal-to-Noise Ratio (PSNR), Structural Similarity Index (SSIM), and Mean Squared Error (MSE) measurements, as well as the best qualitative results when presented to expert radiologists \cite{2024.ROFENA.1,Khorshidifar}. Khorshidifar \textit{et al.} \cite{Khorshidifar} indicates that CycleGAN successfully preserves structural details.  

Hosseinipour \textit{et al.} \cite{Hosseinipour}, applied some U-nets variations and, after realizing that these usually reduce many bright details, proposed the IAMNet model, which yields better performance by effectively highlighting the bright details in the DES image, closely resembling both brightness and location of the bright area found in the ground truth, potentially enough to raise suspicion of potential abnormalities.

After applying other models, including CycleGAN, Rofena \textit{et al.} \cite{rofena2025}, also propose Seg-CycleGAN that synthesizes high-fidelity dual-energy subtracted images from low-energy images, leveraging lesion segmentation maps to guide the generative process and improve lesion reconstruction.  
 
While other studies have developed methods for generating virtual DES images, none have focused on their impact on lesion classification. Instead, these studies have relied on qualitative human evaluation or have simply evaluated the quality of the generated images using metrics such as PSNR, MSE and SSIM \cite{2024.ROFENA.1, Khorshidifar, Hosseinipour}, without considering the impact on the lesion classification system. In this paper, to generate this type of image, we tested several models, namely a pre-trained U-Net model, a U-Net trained end-to-end model, and a CycleGAN model. 
This choice stems from the fact that these have been the most used models and the ones that have provided the best results, so far. To the best of our knowledge, this is the first study to systematically evaluate the impact of virtual DES images on lesion classification performance in CESM. The goal is not to obtain the best classification possible, but to assess the relative difference in performance obtained when using virtual DES images compared to the use of real ones, and to infer if this difference could justify discarding the high-energy image acquisition, potentially reducing radiation exposure.

\section{Methods}
\subsection{Image database}
\label{sec:Database} 

In this work, we used the Categorized Digital Database for Low-Energy and Subtracted Contrast Enhanced Spectral Mammography Images (CDD-CESM) \cite{Khaled2021}, provided by the Radiology Department of the National Cancer Institute of Cairo University, in Egypt. This database contains 2006 low-energy images with the corresponding subtracted CESM images, in both CC and MLO views. Images were collected with informed consent from 326 female patients (aged 18 to 90 years) during the period comprised between January 2019 and February 2021, using two different equipment models: GE Healthcare Senographe DS and Hologic Selenia Dimensions Mammography Systems. Images were acquired in DICOM format. Expert radiologists manually annotated images according to the 2013 American College of Radiology Breast Imaging Reporting and Data System (ACR BIRADS) lexicon, including, among others, breast composition, mass shape, architectural distortion, calcification type, calcification distribution, mass enhancement pattern, and overall BIRADS assessment (1 to 6) \cite{Khaled2022}. Subsequently, the authors of the database manually cropped around 30\% of the images to "remove all irrelevant and unused boundaries". All images were then exported to JPEG format, and abnormality segmentation annotations, verified medical reports, and pathological diagnosis for all cases were made available \cite{Khaled2022}.

\subsection{Image Registration}
\label{sec:Regist}
For some unverified reason, the field of view between the corresponding LE and DES images provided in JPEG format is not the same. For example, image P1\_L\_DM\_MLO (LE) has a matrix size of 1393$\times$2129 pixels, whereas the corresponding CM (DES) image has a matrix size of 1390$\times$2341 pixels). Through visual inspection, we have also verified that the coordinates of the same lesion, visible in both images, did not match. We have also concluded that the localization discrepancies were not affected by rotations, scale changes, or deformations, and that they were not superior to 10 pixels in $x$ or $y$ axes (in both negative and positive directions). Therefore, we have decided to implement a registration program, in Python, aimed at finding the registration parameters (for translations in $x$ and $y$) necessary to bring each pair of images into spatial correspondence. 

The search for the optimal $t_x$ and $t_y$ values was carried out using a two-level search approach: at level 1, the search for the best registration parameters was performed in the $x$ and $y$ directions with a 5-pixel step, over a range between -10 and +10 pixels in each axis, thus totaling 25 search points; in level 2, an exhaustive search was performed in a $5\times5$ pixel neighborhood around the best translation parameters found in level 1. At both levels, the translation parameters (in $x$ and $y$) are always integer values (as the search was conducted taking the pixel as unity, not the pixel size in mm). This was done to avoid interpolation errors when creating the aligned version of the floating image. The measure of similarity used was mutual information \cite{Maes1997, Collignon1995, WELLS1996, viola1997}, where for each image pair, the LE image is the reference one and the DES image is the floating one, given its robustness to the differences in intensity between LE and DES images. The registration result was manually validated by us, through visual inspection.

\subsection{Model}
\label{sec:Model}

Since the idea is to evaluate the impact of using a virtual DES image in a lesion classification scenario, we developed a modular classification architecture that allows experimenting with different virtual DES image generators.

\begin{figure}[!h]
    \centering
    \includegraphics[width=0.9\linewidth]{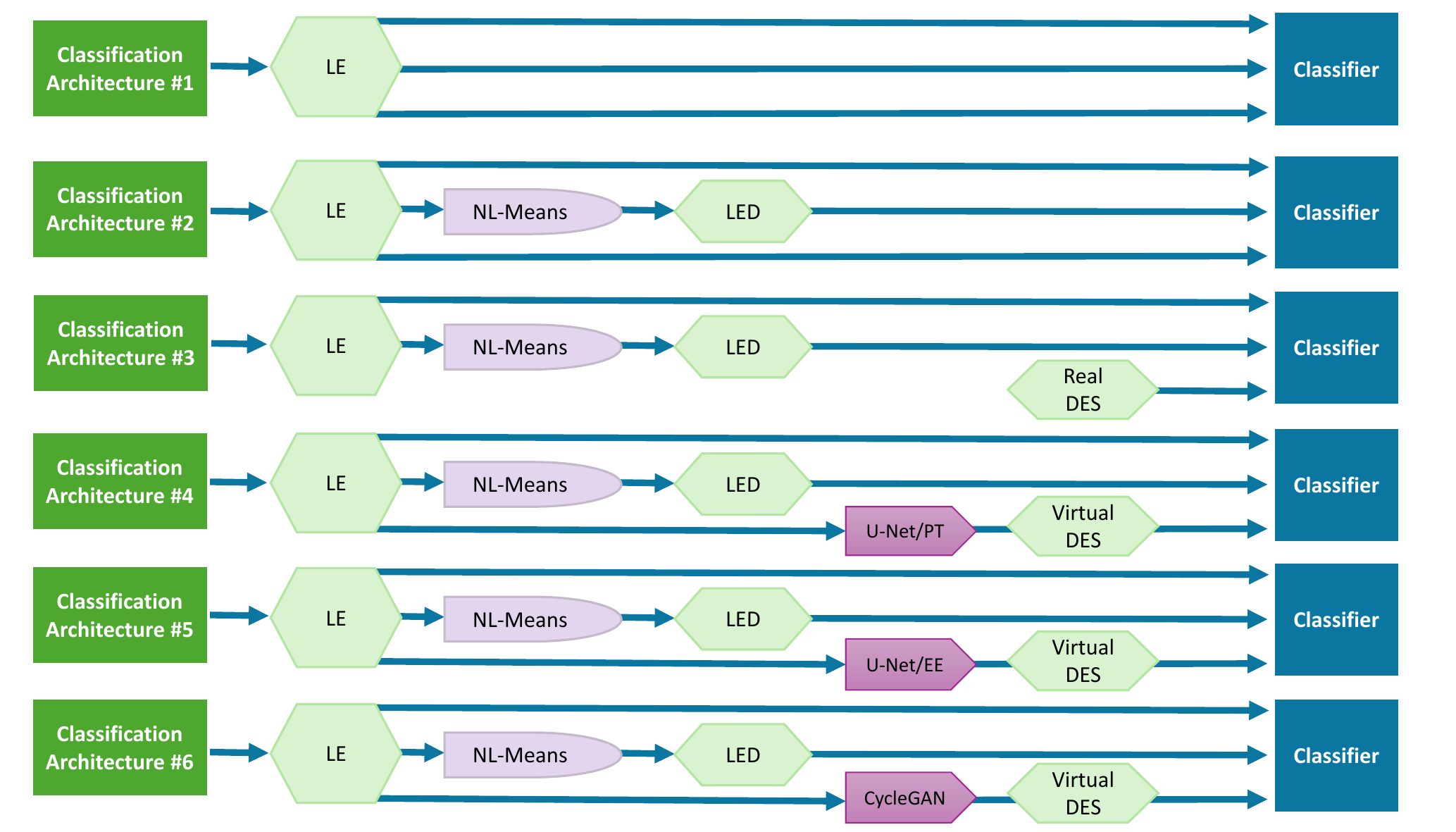}
    \caption{Scheme of the classification architectures studied in this work, numbered from 1 to 6. In the first three classification architectures, the third images given to the classifier are the LE, the LED and the real DES image. In the remaining classification architectures, the third images are virtual DES images produced by the generator models U-Net/PT, U-Net/EE and CycleGAN.}
    \label{fig:architecture1}
\end{figure}

As can be seen in Fig.~\ref{fig:architecture1}, the architecture receives as input the low-energy image (LE) and passes it to the classifier, together with two other images. The second one is a denoised version of the LE image (which we designate as "LED"), obtained by applying the OpenCV non-local means denoising method. The third image received by the classifier is either the real DES image (for a baseline) or a generated image that we wish to use as a possible replacement of the real DES (designated as "virtual DES"). The method used is explained in more detail in Subsection \ref{subsec:inputimages}.  

Like we already mentioned, to generate the virtual DES images, we tested several generator models, namely, a U-Net model either pre-trained or trained end-to-end together with the classifier and a CycleGAN model, which are discussed in the following subsections.

\subsubsection{U-Net}

One way to create a virtual DES image is using a U-Net model \cite{2015.ronneberger.1}. Although it was originally created for image segmentation, a U-Net works for image-to-image translation because it learns a mapping from input to output images, preserves spatial details via skip connections and extracts multi-scale features and reconstructs them in full resolution.
In a first approach, we pre-trained a U-Net to learn to translate from the LE to DES images, using only the training set data. Then we used the trained model to generate virtual DES images for the training, validation and test sets.

A second approach using the U-Net model was made by training it together with the classifier in an end-to-end manner.

\subsubsection{CycleGAN}

In this case, we looked at the state-of-the-art in generation of virtual DES images, and trained the model that was found to have the best performance in \cite{2024.ROFENA.1}. 
The CycleGAN model is designed to learn mappings between two visual domains $X$ and $Y$ without requiring paired examples ($x_i$, $y_i$). Its key innovation is the combination of adversarial training with a cycle‐consistency constraint. 
It uses two generator networks $G: X \rightarrow Y$ and $F: Y \rightarrow X$, each paired with a cor\-res\-pon\-ding discriminator $D_Y$ and $D_X$.  The discriminators are trained to distinguish real images in their domain from "fake" outputs produced by the opposite‐domain generator.

Each pair $(G, D_Y)$ and $(F, D_X)$ is trained with a standard GAN loss, given by
\begin{equation}
L_{GAN}(G, D_Y) = E_y[\log D_Y(y)] + E_x[\log(1 – D_Y(G(x)))],
\end{equation}
\begin{equation}
L_{GAN}(F, D_X) = E_x[\log D_X(x)] + E_y[\log(1 – D_X(F(y)))].\end{equation}

This encourages $G(x)$ to be indistinguishable from real samples in $Y$, and similarly $F(y)$ to mimic the domain $X$.
Because no direct $x \leftrightarrow y$ pairs are needed, it enforces approximate invertibility via
\begin{equation}
L_{cyc}(G,F) = E_x[\parallel F(G(x)) - x \parallel_1] + E_y[ \parallel G(F(y)) - y \parallel_1].
\end{equation}
Minimizing $L_{cyc}$ ensures that translating an image forth and back reconstructs the original content, preventing the generators from arbitrarily mapping all inputs to a single point in the target domain.
An identity mapping loss is also used to ensure that when real samples of the target domain are used in the generator, they are mapped identically to the output. This loss is given by
\begin{equation}
L_{id}(G,F) = E_x[\parallel F(x) – x \parallel_1] + E_y[ \parallel G(y) – y \parallel_1].
\end{equation}
The complete training loss is:
\begin{equation}
L(G,F,D_X,D_Y) = L_{GAN}(G, D_Y) + L_{GAN}(F, D_X) + \lambda_1 L_{cyc}(G,F)+ \lambda_2 L_{id}(G,F),
\end{equation}
where the $\lambda$s balance the trade‐off between visual realism and content preservation.


The CycleGAN model was trained using only the training data set, and it was used afterwards to generate virtual DES images for the training, validation and test sets. These images were then loaded as needed while training and evaluating the classification architecture.
Details on the used training hyperparameters are given below.

\section{Experiments}
\label{sec:experiments}

\subsection{Setup}
\label{subsec:setup}

We first run preliminary experiments using only the training and validation sets. These were meant to assess the best configurations in terms of hyperparameters before we run the final experiments with the test set using only the selected setups.

The preliminary experiments were run for a fixed number of 30 epochs, using learning rates (LR) from the set $\{1\times 10^{-3}, 5\times 10^{-4}, 1 \times 10^{-4}, 5 \times 10^{-5}\}$; the optimizer was Adam and we kept the best validation metrics of the 30 epochs (sufficient for convergence based on validation performance) as the representative of a particular setup. Also, we run all the preliminary experiments for 30 repetitions and present the averages and standard deviations of both the accuracy and the F1 score. Regarding the U-Net model, we varied the number of channels in the first layer ($\{4, 8, 12, 16, 32, 64\}$) and the depth (from 2 to 7). The results showed that the best performance was obtained using 32 channels and depth 6 and, therefore, these values were fixed for the rest of the experiments. We note that better results could probably be obtained using higher values for these parameters, but the resulting generator architectures would not fit in our GPU's RAM. Also, not all combinations of those two parameters could be tested given that the resulting generator architectures would be too large for the GPU. Dropout was tested in the U-Net model and in the classifier with values equal to $\{0.1, 0.2, 0.3\}$. This is important since the size of the data set, even with data augmentation, is not large and the dropout is useful in regularizing the models. We found in the preliminary experiments that using a dropout value of 0.2 was a good setting, hence it was used in the remaining experiments. The used loss function was cross-entropy.

The training parameters for the CycleGAN generator model were: number of epochs = 300, batch size = 8, learning rate = 10$^{-5}$, $\lambda_1$ = 10 and $\lambda_2$ = 5. These values were the ones used originally on \cite{2024.ROFENA.1}, and we did not alter them.

We used a PC with an AMD Ryzen 7 3700X processor, 64 GB RAM, 1TB SSD, with an Nvidia RTX 4090 GPU (24GB of RAM) running Pop!OS 22.04 and Pytorch.

\subsection{The Data Set}
\label{subsec:dataset}

After image registration, $224\times224$ pixel squares were manually cropped from the LE and DES images. For benign and malignant cases, this was done taking into consideration the location of the lesion. For normal cases, this was done randomly within the breast tissue.

\subsubsection{Class Labels}
According to the Breast Imaging Reporting and Data System (BI-RADS) of the American College of Radiology (ACR) \cite{BIRADS}, if the lexicon used in the report corresponds to BI-RADS Category 1 or 2, the mammogram is deemed normal or contains benign findings, so it denotes essentially zero chance of malignancy. These categories do not usually require further imaging and are not expected to change over the follow-up interval. In both, regular screening follow-up is recommended.

Having taken this information into account, we decided to consider two classes: "non-malignant", where we included images considered normal (BI-RADS 1) and benign (BI-RADS 2), and "malignant" (with high suspicious and malignant lesions i.e. BI-RADS $\geq$ 4), where we considered images with malignant lesions. Considering the database used on this work, this decision was supported by the fact that the number of images classified as benign (BI-RADS 2) is small. 

\subsubsection{Data Splits}
\label{subsec:datasplits}

The crops were randomly split into training, validation and test sets such that: 1) crops from a given patient appear in only one of the sets (patient-level splits prevent data leakage), and; 2) the approximate proportion of non-malignant to malignant crops remains similar in the three sets (stratified partition). The proportion of crops in each set is, respectively, 70\%, 15\% and 15\%, for the training, validation and test sets.
The number of images in each split is not perfectly aligned with these percentages given the two restrictions listed, but is very close.
The number of crops and the number of non-malignant and malignant crops per subset is presented in Table \ref{tab:dataset}, along with the actual percentage of images in each set. 

\begin{table}
	\centering
\caption{Data set details: the number of crops per subset, their classes and the percentage of the total number of image crops in each set. \label{tab:dataset}}
\begin{tabular}{lcccc}
	\hline
	Set  & Non-malignant & Malignant & Total crops & Percentage\\
	\hline
	Train  & 561 & 397& 958 & 70.3\\
	Validation  & 104 & 95& 199 & 14.6\\
	Test & 116& 89& 205 & 15.1\\ \hline
	Total & 781&  581&1362 & 100\\
	\hline
\end{tabular}
\end{table}

\subsubsection{Data Augmentation}
\begin{figure}
	\centering
	\includegraphics[width=0.9\linewidth]{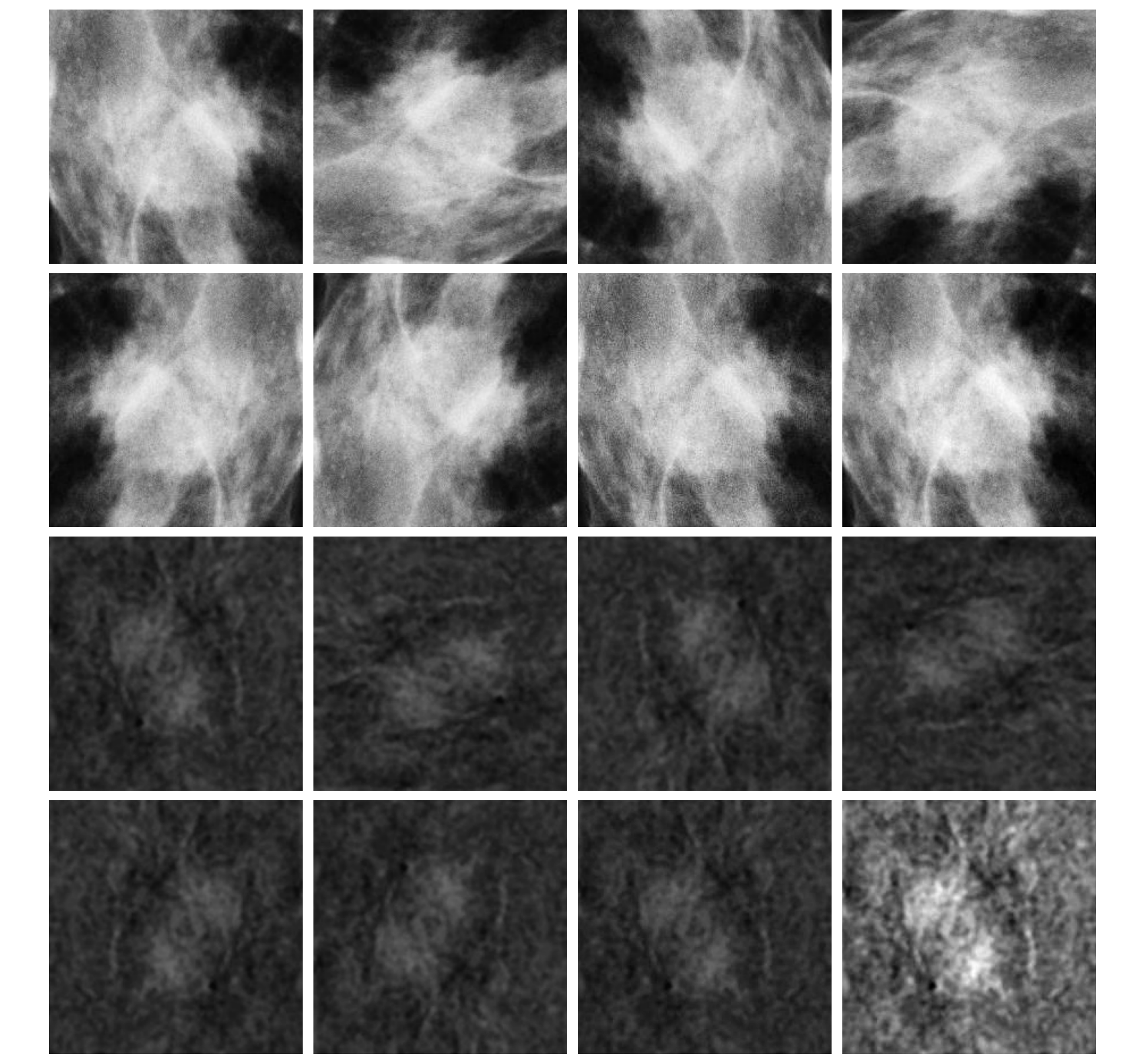}
	\caption{Example of the synchronized data augmentation: both the LE and DES images receive the same random transform every time. Here we exemplify with, from left to right, original input image (LE in the first row, left and DES in the third row, left) then each followed by 90, 180 and 270 rotations, then horizontal and vertical flip and finally sharpness adjustment and autocontrast.}
	\label{fig:dataaugmentation}
\end{figure}

We trained the classifier using data augmentation. The used transformations were horizontal and vertical flips, rotations by multiples of 90 degrees, sharpness adjustment, and autocontrast, preserving the shape of the lesion. These can be applied cumulatively, and care was taken so that when using LE and DES images for training, both images received the same transformation. Figure~\ref{fig:dataaugmentation} contains an example of these transformations. Transformations such as those based on color modifications are not applicable in this context, and changes that can affect the shape of the lesion can negatively influence the results.

\subsubsection{Input Images}
\label{subsec:inputimages}


All experiments used the LE image as the first input image. 
The second image is a denoised version of the LE image (which we designate as "LED"), obtained by applying the OpenCV non-local means denoising method. There is one exception, where we use LE for all three input channels, to serve as a baseline without additional pre-processing or generation (see Fig.\ref{fig:architecture1}, Classification Architecture \#1). The goal of creating the LED image was to remove visible noise in the images, which included JPEG compression artifacts. This method uses three parameters: \emph{h-value}, \emph{templateWindowSize} and \emph{searchWindowSize}. The OpenCV NL-Means implementation we used recommends an \emph{h-value} in the range 10 to 15 for 8-bit medical images; we selected $h$-$value = 10$ because it is conservative, avoiding over-smoothing fine structures while still suppressing blocking artifacts \cite{2005.buades.1}. In the original Non-Local Means paper \cite{2005.buades.1}, patch‐sizes of $5 \times 5$ and $7 \times 7$ are shown to provide a good trade‐off between capturing enough local structure and avoiding over-smoothing.  We used the OpenCV's default choice of a $7 \times 7$ template window that stems directly from those early experiments. We used a search window of $21 \times 21$ pixels that ensures that each reference patch can find good “matches” from a neighborhood large enough to contain structurally similar tissue regions (e.g. nearby glandular patterns) while remaining small enough to exclude distant unrelated anatomy. 

The justification for this choice for the second input image was also supported by the first and second data rows of Table \ref{tab:results1}, for the validation set results, where in the first row we have the results when using only the LE image, and in the second row are the results when the LED was used as the second image: the results when using the LED image improved more than 0.8\% in terms of mean F1 score relative to the use of the LE image.

\begin{figure}[tb] 
  \centering
    \includegraphics[width=0.8\linewidth]{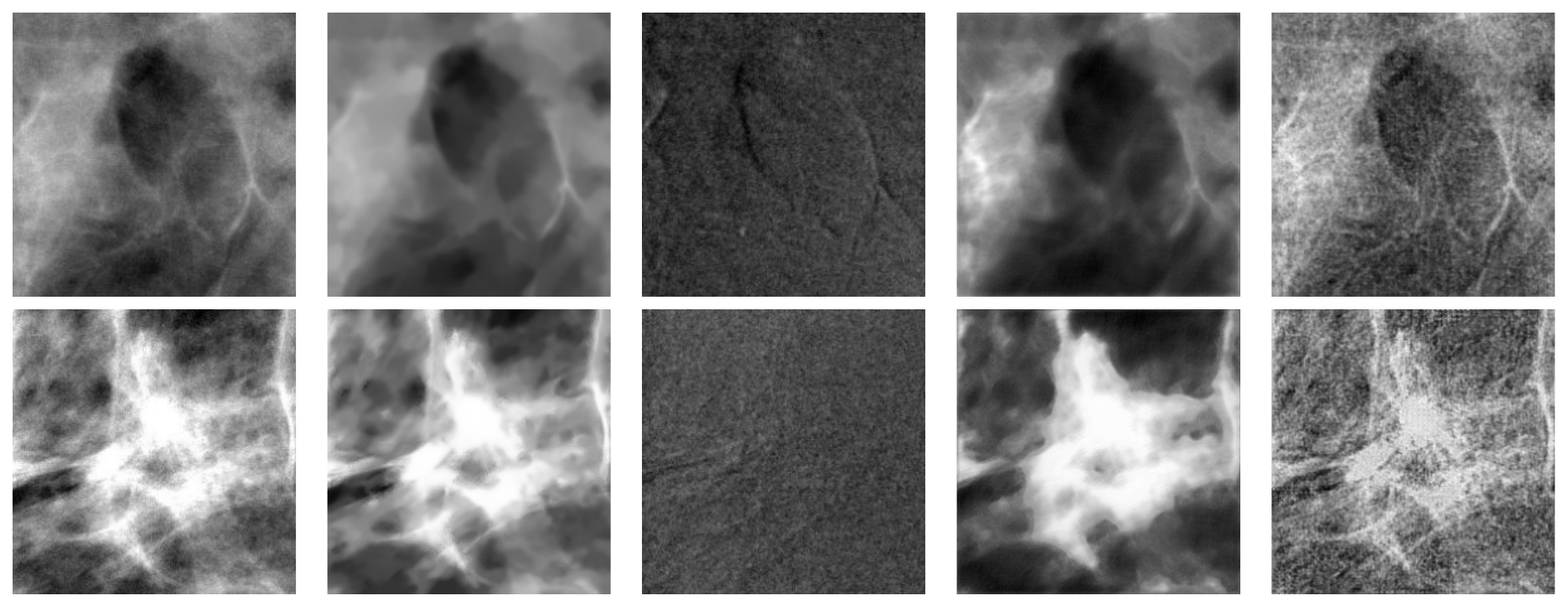}
      \caption{Each row contains, from left to right, one example of LE crop with the respective LED, real DES crop, and the virtual DES images produced using a pre-trained U-Net and the CycleGAN generator architectures.}
  \label{fig:imagens_geradas}
\end{figure}

Regarding the third image, and as baselines, we tested the use of the LE image itself and also the real DES image. These appear on the top part of Table~\ref{tab:results1}, and are in the second and third data rows, respectively.
The other versions will use as the third image, either the output of a U-Net models (pre-trained or trained end-to-end together with the classifier) or the output of a pre-trained CycleGAN model. These options will be called LE+LED+U-Net/PT, LE+LED+U-Net/EE and LE+LED+CG, respectively. An example of images used as input in the classifier is shown in Fig. \ref{fig:imagens_geradas} and, for a better understanding of different classification architectures, see Fig.\ref{fig:architecture1}.  

\subsection{Classifier}
\label{subsec:classifier}

The classifier will be fixed when we vary the types of input images, and as such, it can be considered as a non-fundamental block in the model. The goal is not to obtain the best classification possible, but to assess what is the relative difference in performance that is obtained when using the different input images, to understand if the use of virtual DES images is justifiable. Nonetheless, we tested three difference networks to serve as the classifier: two variants of a ResNet \cite{resnet}, namely ResNet V1.5 in the form of a ResNet18, and a ResNet34, and also a RegNetY\_1.6GF \cite{Radosavovic_2020_CVPR}. After several preliminary experiments, the latter consistently achieved the best results, hence we fixed it as the classifier for the remaining experiments. These experiments appear in Table \ref{tab:classifier} and were also used to determine the best learning rate by varying it along with the classifiers, with values in of 0.001, 0.0005 and 0.0001. The chosen learning rate was 0.0005. The input images were three identical LE images, corresponding to the classification architecture \#1. These experiments were conducted using only the train and validation sets.

\begin{table}[!htb]
\centering
\caption{Mean and standard deviation in percentage, for 30 repetitions of the accuracy and F1 score for different classifiers and learning rates. The input images were all LE, corresponding to the classification architecture \#1. Results on the validation set. \label{tab:classifier}}
\begin{tabular}{lrccc}
	\hline
\multirow{2}{*}{Classifier} & Learning &  Mean Valid. & Mean Valid. \\
	& rate &  Accuracy (std) & F1 (std) \\ \hline
ResNet18 & 0.0001& 86.77	(1.88)&	85.53	(2.44) \\ 
ResNet18 & 0.0005& 87.99	(1.90)&	87.41	(2.21) \\ 
ResNet18 & 0.001&  86.21	(1.58)&	85.98	(1.72)\\ 
ResNet34 & 0.0001& 88.21	(1.68)&	87.21	(2.04) \\ 
ResNet34 & 0.0005& 88.06	(1.33)&	87.64	(1.47) \\ 
ResNet34 & 0.001& 86.05	(1.46)&	85.51	(1.84) \\ 
RegNetY\_1.6GF & 0.0001& 87.76	(1.64)&	86.47	(1.99) \\ 
RegNetY\_1.6GF & 0.0005& 89.01	(1.78)&	88.27	(2.06) \\ 
RegNetY\_1.6GF & 0.001& 88.51	(1.99)&	87.72	(2.47) \\ 
\hline
\end{tabular}
\end{table}

\section{Results and Discussion}
\label{sec:ResDisc}

The data we are using are not balanced: there are 781 images from the \emph{Non-malignant} class and 581 from the \emph{Malignant} class, as discussed above.
Although we show both accuracy and F1 score in Table \ref{tab:results1}, we argue that the F1 score is the relevant metric, since in imbalanced classification scenarios accuracy can be misleading because a classifier might simply favor predicting the majority class and still achieve high accuracy overall. The F1 score, on the other hand, combines precision and recall into a single metric, thus providing a better understanding of how well the classifier performs in each class, especially on the minority ones. This makes F1 a more informative and fair metric when we care about detecting the minority class correctly. Additionally, the F1 score is particularly useful when the costs of false negatives and false positives are high, such as in medical diagnosis, which is our case. 

The results presented in Table~\ref{tab:results1} show the performance for different input image configurations in the 6 classification architectures. The evaluation metrics, accuracy and F1 score, were computed over 30 repetitions on both validation and test sets.

\begin{table}[tb]
\centering
\caption{Mean and standard deviation in percentage, for 30 repetitions of the accuracy and F1 score for the 6 classification architectures with different input images and generator models on the validation and test sets. \label{tab:results1}}
\begin{tabular}{lccccc}
	\hline
\multirow{2}{*}{Classification model (input images)} & \multicolumn{2}{c}{Validation set} &  & \multicolumn{2}{c}{Test set} \\
\cline{2-3} \cline{5-6}
& Mean Acc. (std) &  Mean F1 (std) &  & Mean Acc. (std)  & Mean F1 (std) \\
\hline
\#1 (LE+LE+LE) & 89.01	(1.78)	&88.27	(2.06) &  &	86.13	(2.22)	& 85.38	(2.07) \\ 
\#2 (LE+LED+LE) & 89.65	(1.57)	&89.10	(1.70) &  & 85.24	(2.00) &	84.67	(1.97) \\ 
\#3 (LE+LED+DES) & 91.21	(1.65)	&90.62	(1.77) &  &91.48	(1.52)	& 90.35	(1.72)\\ 
\hline
\#4 (LE+LED+U-Net/PT) & 90.13	(1.89) & 89.58	(2.07) &  & 86.26	(2.18) &	85.59	(2.06) \\ 
\#5 (LE+LED+U-Net/EE) & 89.51	(1.82)&	89.02	(2.07) &  & 86.15	(2.47) & 85.43	(2.37) \\ 
\#6 (LE+LED+CG)	& 88.86	(2.02)&	88.20	(2.25) &  & 84.91	(2.04)	& 84.30	(1.86) \\ 
	\hline
\end{tabular}
\end{table}

The baseline classification models (first three data rows) provide a reference for evaluating the virtual DES approaches. The first classification architecture \#1 (with images LE+LE+LE) uses three low-emission images and achieves a mean validation accuracy of 89.01\% (std: 1.78\%) and an F1 score of 88.27\% (std: 2.06\%), with test set performance slightly lower at 86.13\% accuracy (std: 2.22\%) and 85.38\% F1 score (std: 2.07\%). 

Incorporating a denoised low-emission image (Classification architecture \#2 with images LE+LED+LE) slightly improves validation performance to 89.65\% accuracy (std: 1.57\%) and 89.10\% F1 score (std: 1.70\%), but test set performance remains comparable to the LE+LE+LE configuration (85.24\% accuracy, std: 2.00\%; 84.67\% F1 score, std: 1.97\%). This suggests that denoising the LE image provides marginal benefits, likely by reducing artifacts such as JPEG compression noise, but does not substantially enhance diagnostic performance.

The third classification architecture \#3 (with images LE+LED+DES) incorporates a real DES image and demonstrates the highest performance across all classification architectures, with a mean validation accuracy of 91.21\% (std: 1.65\%) and F1 score of 90.62\% (std: 1.77\%), and test set results of 91.48\% accuracy (std: 1.52\%) and 90.35\% F1 score (std: 1.72\%). This superior performance highlights the diagnostic value of real DES images, which leverage the contrast enhancement from dual-energy imaging to improve the detection of malignant features. 

The virtual DES configurations in the second half of the table (last three rows) aim to replicate the diagnostic utility of real DES images while minimizing radiation exposure. The classification architecture \#4, with a LE+LED+U-Net/PT configuration, where a U-Net model is trained and generates the virtual DES images before the classifier is trained, achieves a mean validation accuracy of 90.13\% (std: 1.89\%) and an F1 score of 89.58\% (std: 2.07\%), with test set performance of 86.26\% accuracy (std: 2.18\%) and 85.59\% F1 score (std: 2.06\%). These results are competitive with the classification architecture \#2 (LE+LED+LE baseline) and approach the performance of the classification architecture \#3 (real DES), particularly on the validation set. The pre-trained U-Net generator produces virtual DES images that retain critical diagnostic information.

The LE+LED+U-Net/EE configuration, in the classification architecture \#5, where the U-Net model is trained end-to-end with the classifier, yields slightly lower performance, with a validation accuracy of 89.51\% (std: 1.82\%) and F1 score of 89.02\% (std: 2.07\%), and test set results of 86.15\% accuracy (std: 2.47\%) and 85.43\% F1 score (std: 2.37\%). The end-to-end training approach may suffer from optimization challenges, as the joint training of the generator and classifier could lead to suboptimal convergence compared to the pre-trained U-Net model. However, the test set F1 performance remains above the LE+LED+LE baseline \#2, indicating that these virtual DES images still provide meaningful diagnostic information.

The LE+LED+CG architecture \#6, using a pre-trained CycleGAN model to generate virtual DES images, performs the least effectively among the virtual DES approaches, with a validation accuracy of 88.86\% (std: 2.02\%) and F1 score of 88.20\% (std: 2.25\%), and test set results of 84.91\% accuracy (std: 2.04\%) and 84.30\% F1 score (std: 1.86\%). The lower performance of CycleGAN model may be attributed to its adversarial training framework, which prioritizes visual realism over diagnostic fidelity, potentially introducing artifacts or losing subtle features critical for malignancy detection.
The degradation of around 6\% in F1 score in the test set relative to the LE+LED+DES baseline indicates that the CycleGAN's mapping does not preserve the subtle contrast features that the detector uses for malignancy discrimination. The higher standard 
deviations also suggest that the CycleGAN introduces additional variability that is not present when the real DES image is supplied.

\section{Conclusion}
\label{sec:Conc}

The idea of producing virtual DES images from the LE images is tempting and has been pursued in the past \cite{2024.ROFENA.1, rofena2025, Hosseinipour, Khorshidifar}. In this scenario, the high-energy image would no longer be acquired, which would reduce the patient's exposure to radiation.

This work is the first to quantify the classification performance gap between virtual and real DES images in CESM, providing a foundation for future research.
We studied if such images can be a valid replacement for the real DES images when it comes to training a classification model to differentiate malignant and non-malignant breast images. 
To achieve this, we first registered the LE and DES images from an open-source CESM database \cite{Khaled2021}, after that, crops were manually produced from selected images and we created a general architecture that allowed us to compare the classification performance between the use of real and virtual DES images. Finally, we conducted several experimental tests to understand the impact of using the virtual DES images.

We found that, as expected, the real DES images contain additional information that helps to achieve higher F1 scores. The difference in performance for our setup shows that the use of virtual DES images does not match the classification quality obtained with the real DES images. 

The work presented has some limitations, namely, 1) the number of CESM images available is small and does not provide an adequate setting for training larger models, despite the use of data augmentation; 2) the fact that we proposed a specific pipeline using three input images to our classifier, although it is possible to devise different approaches; and 3) the data set providers used JPEG compression which introduces undesirable compression artifacts in the images (a format like DICOM would be desirable); 4) other virtual DES approaches could be used and influence the conclusions obtained here.

Possible follow-up work could focus on creating improved virtual DES image generation methods by incorporating domain-specific knowledge (e.g., vascularity patterns), using larger data sets, or exploring other generative models like diffusion models, before a risk‑benefit assessment can justify the elimination of real DES images' acquisition.

\funding{This work is supported by UID/04516/NOVA Laboratory for Computer Science and Informatics (NOVA LINCS) with the financial support of FCT.IP.}

\roles{
Ana Perre: Preparation, creation and/or presentation of the published work, specifically writing the initial draft; Data curation, writing—review and editing. Lu\'{i}s Alexandre: Programming, software development; designing computer programs; implementation of the computer code and supporting algorithms; testing of existing code components; writing—review and editing.  Lu\'{i}s Freire: Programming, software development; writing—review and editing, specifically critical review, commentary or revision.}


\data{The data set used was obtained from - www.cancerimagingarchive.net/collection/cdd-cesm/}


\bibliographystyle{plain}
\bibliography{Bibliography}

\end{document}